\documentstyle[11pt,aas2pp4]{article}
\received{20 May 20 1998}
\accepted{30 March 1999}
\slugcomment{To appear in ApJ}
\lefthead{Hirayama et al.}
\righthead{Observations of PSR~B1259$-$63 System}
\newcommand{\psr}{PSR~B1259$-$63}
\newcommand{\system}{PSR~B1259$-$63/SS~2883 system}
\newcommand{\ser}{AX~J1302$-$64}
\newcommand{\msol}{\mbox{{\it M}$_{\odot}$}}
\newcommand{\rsol}{\mbox{{\it R}$_{\odot}$}}
\begin{document}

\title{X-Ray/Gamma-Ray Observations of the \system\ near Apastron}

\author{M. Hirayama\altaffilmark{1}}
\affil{Santa Cruz Institute for Particle Physics,
University of California, Santa Cruz, Santa Cruz, CA 95064;
hirayama@scipp.ucsc.edu}
\altaffiltext{1}{Research Fellow,
the Japan Society for the Promotion of Science for Young Scientists}

\author{L. R. Cominsky}
\affil{Department of Physics and Astronomy, Sonoma State University,
Rohnert Park, CA 94928;lynnc@charmian.sonoma.edu}

\author{V. M. Kaspi}
\affil{Center for Space Research and Department of Physics,
Massachusetts Institute of Technology, Cambridge, MA 02139;
vicky@space.mit.edu}

\author{F. Nagase}
\affil{The Institute of Space and Astronautical Science,
1-1 Yoshinodai 3-chome, Sagamihara, Kanagawa 229, Japan;
nagase@astro.isas.ac.jp}

\author{M. Tavani}
\affil{Columbia Astrophysics Laboratory, New York, NY 10027\\
IFCTR-CNR, via Bassini 15, I-20133 Milano, Italy; tavani@astro.columbia.edu}

\author{N. Kawai}
\affil{The Institute of Physical and Chemical Research,
2-1 Hirosawa, Wako, Saitama 351-01, Japan;
kawai@postman.riken.go.jp}

\and

\author{J. E. Grove}
\affil{E. O. Hulburt Center for Space Research,
U. S. Naval Research Laboratory, MS 7650, Washington, D. C. 20375;
grove@osseq.nrl.navy.mil}

\begin{abstract}
We report the results from X-ray and hard X-ray observations of the
\system\ with the {\it ASCA} and {\it CGRO} satellites, performed
between 1995 February and 1996 January when the pulsar was near
apastron.  The system was clearly detected in each of the two {\it
ASCA} observations with luminosity in the 1--10~keV band of $L_{\rm X}
= (9 \pm 3) \times 10^{32}\ (d/2{\rm ~kpc})^2$~erg~s$^{-1}$, while CGRO/OSSE
did not detect significant hard X-rays from the system.  X-ray spectra
obtained with {\it ASCA} are well-fit by a single power-law spectrum with a
photon index of $1.6 \pm 0.3$.  No pulsations were detected in either
the {\it ASCA} or the OSSE data.  We combine all existing X-ray and
hard X-ray observations and present the orbital modulation in the
luminosity and photon index for the entire orbit.  The results are in
agreement with predictions based on a synchrotron emission model from
relativistic particles in a shocked pulsar wind interacting with the
gaseous outflow from the Be star.
\end{abstract}


\section{Introduction}

The \system\ is a binary consisting of the radio pulsar \psr\ and the Be
star SS~2883.  \psr\ is a 48~ms radio pulsar, discovered by Johnston et
al. (1992a).  The pulsar's parameters were determined by Manchester et
al. (1995) using radio timing observations with the Parkes 64-m radio
telescope, covering the two periastron passages in 1990 and in 1994.
The radio timing observations have shown that the pulsar is in a 3.4~yr,
binary orbit with eccentricity $e = 0.86$.  Further radio observations
with the Parkes telescope suggest that the binary orbit may be
precessing by the quadrupole moment of the tilted companion star (Wex et
al. 1998).  The companion star has been identified as the 10th magnitude
B2e star SS~2883 from optical observations, and the mass and radius of
SS~2883 were estimated to be $\sim 10$~\msol\ and $\sim 6$~\rsol,
respectively (Johnston et al. 1992b).

 From the dispersion measure (DM) of the pulsar and a model for the
galactic electron distribution (Taylor \& Cordes 1993), the
estimated distance to the source is 4.6~kpc, although DM-derived
distances are uncertain often to a factor of $\sim$2.  Johnston et
al. (1994) argue on the basis of optical photometric observations
that the distance to SS~2883 cannot be greater than 1.5~kpc.  In
this paper we adopt a compromise distance of 2~kpc.

The first attempt to detect X-rays from the \system\ was in
observations made using the {\it Ginga} satellite in 1991.
However, {\it Ginga} failed to detect significant X-ray emission
from the system (Makino \& Aoki 1994).  The first low-level X-ray
detection was by Cominsky, Roberts, \& Johnston (1994) who
observed the system just after apastron using {\it ROSAT} in
September 1992 (``{\it ROSAT} obs~2'' and ``{\it ROSAT} obs~3'' in
Fig.~\ref{figure:psr-orbit}).  The X-ray luminosity in the
1~--~10~keV band during these observations can be derived from the
{\it ROSAT} results as (0.07~--~2) $\times 10^{33} (d/2{\rm ~
kpc})^2$~erg~s$^{-1}$ for {\it ROSAT} obs~2 and (0.3~--~1.5)
$\times 10^{33} (d/2{\rm ~ kpc})^2$~erg~s$^{-1}$ for {\it ROSAT}
obs~3, respectively, assuming the spectral parameters given by
Cominsky, Roberts, \& Johnston (1994).  A subsequent analysis by
Greiner, Tavani, \& Belloni (1995) of archival {\it ROSAT} data
taken just before apastron in February 1992 (``{\it ROSAT} obs~1''
in Fig.~\ref{figure:psr-orbit}), reveals significant X-ray
emission at level consistent with the result of Cominsky et~al.
(1994).

The \system\ was observed with the {\it ASCA} satellite at six
different orbital positions (hereafter {\it ASCA} obs~1 through
{\it ASCA} obs~6) and with the OSSE instrument on-board the {\it
CGRO} satellite at two different orbital positions: one near
periastron and the second near apastron (hereafter OSSE obs~A and
OSSE obs~B).  {\it ASCA} observations 1--3 have been previously
reported by Kaspi et al. (1995), {\it ASCA} obs~4 has been
reported by Hirayama et al. (1996), and OSSE obs~A has been
discussed by Grove et al. (1995).  In
Figure~\ref{figure:psr-orbit} we provide a schematic drawing of
the pulsar's orbit around the systemic center of mass, together
with the approximate location of the pulsar during all six {\it
ASCA} observations, the three {\it ROSAT} observations reported by
Cominsky et al. (1994) and Greiner et al. (1995), and the two OSSE
observations.  Also, for completeness,
Table~\ref{table:observation} gives a summary of the six {\it
ASCA} and two OSSE observations and of the orbital
geometry for an assumed Be star mass $M_c = 10$~\msol.

In this paper, we present the new {\it ASCA} data taken in {\it
ASCA} obs~5 and {\it ASCA} obs~6, which were obtained near
apastron, at phases similar to the {\it ROSAT} observations which
occurred one 3.4 yr orbital cycle earlier (Cominsky et al. 1994;
Greiner et al. 1995).  We also present the OSSE obs~B apastron
results.  The observations provide the X-ray luminosity in the
2~--~10~keV band at apastron which places constraints on the
theoretical models (e.g. Tavani, Arons, \& Kaspi 1994; Tavani \&
Arons 1997). Also, using the apastron observations we can attain
maximum sensitivity to pulsed magnetospheric X-ray emission from
the pulsar, as X-ray emission from the binary interaction is at a
minimum.

\section{Observations}

The {\it ASCA} satellite (Tanaka et al. 1994) carries four X-ray
telescopes (Serlemitsos et al. 1995), with two X-ray CCD cameras
(SIS hereafter) at two of their focal planes, and two gas
scintillation imaging proportional counters (GIS hereafter, Ohashi
et al. 1996) at the other two. The SIS detectors cover the energy
range from 0.4~keV to 12~keV with an $11' \times 11'$ square field
of view in 1-CCD mode, and the GIS detectors cover from 0.7~keV to
15~keV with a circular field of view of $25'$ radius.  As with the
four earlier {\it ASCA} observations, during {\it ASCA} obs~5 and
{\it ASCA} obs~6, the two SIS detectors were operated in 1-CCD
faint mode with a time resolution of 4~s.  {\it ASCA}'s two GIS
detectors were in PH mode with time resolution better than 4~ms,
high enough to potentially study the 48~ms pulsations from \psr.

The OSSE instrument consists of four large-area NaI(Tl)--CsI(Na)
phoswich detector systems (Johnson et al. 1993) and covers the
energy range from 50~keV to 10~MeV with good spectral resolution,
and a field of view of approximately $3^{\circ}.8 \times 11^{\circ}.4$.  
OSSE had previously observed this object in 1994 January at periastron.
In both observations, care was taken to minimize the effect of the
galactic diffuse continuum emission and to avoid nearby bright
point sources, such as the X-ray binaries GX~301--2, Cen~X-3, and
2S~1417--624.  The observations have similar orientations with
respect to the galactic plane and similar background fields of
view, and should therefore be subject to similar, minimal
systematic effects.  For the two week OSSE observation, spectra
were accumulated in a sequence of two-minute measurements of the
source field alternated with two-minute, offset-pointed
measurements of background fields.  In a total exposure time of $4.8
\times 10^5$ s, the highest-quality data were collected on the
source field, with an approximately equal time on the background
fields.  Simultaneously with the spectroscopy data, count-rate
samples in eight energy bands at 8-ms resolution were collected to
search for pulsed emission at the radio period.

One possible systematic effect is the inclusion of flux from the
serendipitous source \ser, in the OSSE data.  This weak source is seen
in both the {\it ROSAT} data (Greiner et. al. 1995) and the {\it ASCA}
data (Kaspi et al. 1995; Hirayama et al. 1996), but does not
contaminate either analysis due to the imaging nature of the
instruments.  It is possible, however, that the source has flux in the
OSSE bandpass, which could have added to that detected in this region
in the periastron OSSE obs~A data, although the required spectrum
would be unusual for an accreting binary.  We argue that the
correlated variability seen in the {\it ASCA} and OSSE observations,
in which the flux at periastron is more than a factor of ten greater
than that detected near apastron in both instruments, makes it
unlikely (but not impossible) that the OSSE obs~A flux arises from
\ser. A detailed analysis of \ser\ will be presented elsewhere.

\section{Data Analysis \& Results}

\subsection{Spectral Results from ASCA data}

In both observations {\it ASCA} obs~5 and {\it ASCA} obs~6, the
\system\ was clearly detected with both SIS and GIS detectors.
The observed positions of the X-ray source were coincident with
the position of \psr\ to within $1'$, the accuracy of the attitude
determination of the {\it ASCA} satellite.  The method of analysis
used to reduce the data from {\it ASCA} obs~5 and {\it ASCA} obs~6
is identical to that used for the first four {\it ASCA}
observations (Kaspi et al. 1995; Hirayama et al. 1996).

After subtracting the background spectra, the source spectra from
the combined {\it ASCA} GIS and SIS data taken in {\it ASCA} obs~5
and {\it ASCA} obs~6 were fitted with a single power law model
with photoelectric absorption, using XSPEC, a software package for
X-ray spectral analyses.  For completeness, the results of the
independent three-parameter fits with the single power law model
are given in Table~\ref{table:simultaneous-fit} for all six {\it
ASCA} observations.  We also fitted the data with a thermal
bremsstrahlung model with photoelectric absorption.  In all cases,
reduced $\chi^2$-values were only slightly higher for the thermal
model compared with that for the power-law model, and we could not
rule out the former on the basis of the {\it ASCA} observations
alone.  No evidence was found for any line features in the
spectra.  The upper limits for Iron K emission line at 6.4~keV,
6.7~keV, or 6.9~keV flux are listed in
Table~\ref{table:simultaneous-fit}.  The spectral fits with a
thermal bremsstrahlung model result in plasma temperatures of
6~--~14~keV, at which Iron K emission lines are expected. Since
none were detected, the absence of these lines argues in favor of
the power law model.

Due to the weak detections at apastron, the X-ray photon indices and
column densities in Table~\ref{table:simultaneous-fit} are consistent
with the entire range of values observed at periastron. In
Figures~\ref{figure:index-vs-separation} and
\ref{figure:flux-vs-separation} the changes in the photon index and the
flux are plotted against the binary separation, assuming masses of
1.4~\msol\ for \psr\ and 10~\msol\ for SS~2883. The X-ray flux at
apastron is $\sim10$ times smaller than that near periastron.  In this
figure, we also show luminosity results from the previous X-ray missions
(Makino \& Aoki 1994; Cominsky et al. 1994; Greiner et al. 1995)
converted into the 1--10~keV band, taking into account the best-fit
photon indices and their errors.

\subsection{Search for X-ray Pulsations in the ASCA data}

We have searched the obs~5 and obs~6 data for evidence that the
X-ray emission from the \system\ is pulsed at the 48~ms radio
period using the same procedures followed by Kaspi et al. (1995)
and Hirayama et al. (1996). These procedures were: 1) epoch
folding; 2) $Z^2_n$ test (Buccheri et al. 1983); 3) search in
$P-\dot{P}$ space (Kaspi et al. 1995).  The $Z^2_n$ test is used
to detect the presence of up to $n$-th harmonic components of the
pulsations (See Buccheri et al. 1983 for details). In the
$P-\dot{P}$ search, pulsations are sought over a certain range of
pulse period $P$ and period derivative $\dot{P}$, around the $P$
and $\dot{P}$ estimated by radio timing observations (See \S 3.2.2
in Kaspi et al. 1995 for details).

The first two searches were done within $\pm 1\mu$s from the
expected pulse period of \psr, as determined by Manchester et al.
(1995) with the radio observations.  Even though Wex et al. (1998)
showed that the timing model for \psr\ by Manchester et al. (1995)
does not fit the radio data obtained after MJD 49600, the period
range we chose was wide enough to cover the entire range of
possible periods.  Searches for pulsations using the epoch-folding
method were also done with a varying number of phase bins (8, 16,
and 32) in order to ensure sensitivity to a variety of pulse
profiles. The $Z^2_n$ tests were performed for the harmonic
numbers $n = 1$, 2, 3, and 4. In addition, searches in the
0.5--2~keV and 2--10~keV bands were performed individually to
detect possible narrow-band pulsations.

In conclusion, we detected no X-ray pulsations at the radio period,
using any pulse searching technique.  The upper limits to the pulsed
component were estimated using the epoch-folding method described in
Leahy et al. (1983) and are listed in
Table~\ref{table:upper-limit}.  The limits were estimated by the
pulsation searches with the epoch-folding method with $j=32$ for three
energy bands: 0.5--10~keV (full band), 0.5--2~keV, and 2--10~keV.  The
$Z^2$ searches (Buccheri et al. 1983) yielded similar results.  In the
table, upper limits on the amplitude of counting rates, $A N_{\gamma}$
divided by net exposure time, are also listed.

\subsection{OSSE Data Analysis and Results}

The method of analysis was similar to that used for the periastron data,
and is described more fully by Grove et al. (1995).  No statistically
significant unpulsed emission is detected by OSSE from PSR~B1259--63
near apastron.  Upper limits in several energy bins spanning 50 keV to
10 MeV are shown in Figure~ \ref{figure:wideband-spectrum}, together
with the {\it ASCA} binned data for both obs~5 and obs~6. Also shown for
comparison are the {\it ASCA} obs~1--4 and OSSE obs~A periastron
spectra, which have fluxes about ten times greater.  At periastron, OSSE
detected emission between 50 keV and $\simeq$200 keV at
$\simeq$5$\sigma$ significance, as reported by Grove et al. (1995).
Based on nearby measurements of the diffuse emission and on galactic
symmetry, Grove et al.  argued that the positive detection at periastron
was probably not due to residual galactic emission.  The null detection
we report here strengthens this argument, and severely restricts the
possibility that small-scale, local fluctuations in the diffuse emission
could have been responsible for the observed flux at periastron.

\section{Discussion}

Our results from studying the soft and hard X-ray emission from
the \system\ can be summarized as follows: 1) the X-ray emission
is non-thermal and unpulsed for entire orbit (Kaspi et al. 1995;
Hirayama et al. 1996; this paper), and especially at periastron, is
characterized by a power-law spectrum which extends from 1~keV to
up to 200~keV with a single photon index of $\approx 2$ (Grove et
al. 1995); 2) the photon index varies slightly with orbital phase
(Fig.~2); 3) the X-ray luminosity in the 1~--~10~keV band varies
with orbital phase by about an order of magnitude, from $\sim
10^{34}$~erg~s$^{-1}$ at periastron to $\sim 10^{33}$~erg~s$^{-1}$
at apastron (assuming that the distance to the system is 2~kpc)
(Fig.~3);  4) the observed X-ray luminosity at the apastron
passage in 1995 (this paper) is consistent with that at the
previous apastron passage in 1992 (Cominsky et al. 1994; Greiner
et al. 1995).  In addition, a recent analysis of data obtained with 
the ASM instrument on the {\it RXTE} satellite near the 1997 periastron 
passage yielded an upper limit to the X-ray flux of $(0.5 \pm 2.4)
\times 10^{-11}$~erg~s$^{-1}$~cm$^{-2}$ in the 2~--~10~keV band
(Kaspi \& Remillard 1998), which is also consistent with the {\it
ASCA} results at periastron by Kaspi et al. (1995).

The consistencies in X-ray luminosity at the two apastron passages
({\it ROSAT} obs~1, 2, and 3 versus {\it ASCA} obs~5 and 6) and at
the two periastron passages ({\it ASCA} obs~2 versus {\it
RXTE}/ASM) support the idea that the observed time variability in
the X-ray luminosity is a binary modulation, i.e., variation due
to the change in the pulsar's position in the binary system.
Therefore, a model in which the X-ray emission arises from only
one of the binary components, such as emission from cooling
neutron star surface, is unlikely to explain the observed data.
Instead, a model that invokes interaction between the pulsar and
the Be star is more likely to explain the observed orbital
phase-dependent time variability in X-ray luminosity.

Binary modulation of the X-ray luminosity and the observed photon
index can quantitatively be interpreted with the non-thermal
diffuse nebular emission model of Tavani \& Arons (1997).  In this
model, the observed X-rays are due to synchrotron emission from
$e^\pm$ pairs with $\gamma = 10^6$ -- $10^7$ accelerated at a
shock front formed between the pulsar wind from PSR~B1259$-$63 and
the stellar wind from SS~2883 (Tavani et al. 1994; Kaspi et al.
1995; Hirayama et al. 1996; Tavani \& Arons 1997). Alternative
explanations of the X-ray emission from the binary system have
also been proposed, such as emission from accretion onto neutron
star surface and emission from material captured outside the
pulsar's light-cylinder (King \& Cominsky et al. 1994).  These
mechanisms predict thermal bremsstrahlung emission from gaseous
material heated through the accretion process, which is less
likely to be consistent with our data.

Results from the {\it ASCA} and OSSE observations near periastron
in 1994 (Kaspi et al. 1995; Grove et al. 1995) indicate that a
shock-powered emission model provides a natural way to account for
all the observations (Tavani et al. 1994; Tavani \& Arons 1997).
Tavani \& Arons (1997) showed that shock-powered emission with
radiative cooling explains the X-ray properties observed around
periastron, such as the power-law spectrum extending to 200~keV,
the spectral softening at periastron, and the luminosity variation
near periastron.  In addition, the results of {\it ASCA} obs~4
through obs~6 (Hirayama et al. 1996; this paper) show qualitative
agreement with the predictions of this model for the X-ray
luminosity and photon index.  On the other hand, accretion-powered
emission models cannot explain the moderate X-ray luminosity of
$\sim 10^{34}$~erg~s$^{-1}$ at periastron, because the strong
magnetic field of the pulsar will inhibit accretion until the mass
flux is sufficiently strong to overcome the centrifugal barrier
(Stella, White and Rosner 1986). Once accretion occurs, a
significantly larger X-ray luminosity should be observed ($\sim
10^{37}$~erg~s$^{-1}$). Also, as Grove et al. (1995) discussed, a
single power-law spectrum from 1~keV to 200~keV cannot be
explained by thermal emission from heated accreting matter.

Based on the shock emission model by Tavani \& Arons (1997), the
power-law spectrum with photon index $\alpha \approx 1.6$,
observed when the pulsar was farther than 1000~lt-s from the Be
star, required the shock acceleration mechanism to create $e^\pm$
pairs with an energy spectrum $N(\gamma) \propto \gamma^{-2}$ just
behind the shock front (Hirayama et al. 1996).  In this context,
the apparent softening of the spectral index can be understood in
terms of enhanced radiative cooling (Kaspi et~al. 1995; Hirayama
et al. 1996; Tavani \& Arons 1997).  The decrease in X-ray flux
near periastron provides additional evidence in favor of this
explanation.

The {\it ASCA} obs~5 data set the tightest upper limit yet on
pulsed X-ray magnetospheric X-ray emission from PSR~B1259$-$63. From
Table~\ref{table:upper-limit}, the pulsed luminosity is $< 3
\times 10^{31}$~erg~s$^{-1}$ at the 99\% confidence level,
assuming a distance of 2~kpc and 1~sr beaming angle.  Saito (1997)
and Saito et al. (1998) systematically studied the pulsed
luminosity from spin-powered pulsars in the X-ray band based on
the {\it ASCA} observations and found an empirical relationship
between pulsed luminosity and spin-down luminosity given by
$L_{\rm X(pulsed)} = 10^{34} \times (\dot{E}_{\rm
rot,38})^{3/2}$~erg~s$^{-1}$, where $L_{\rm X(pulsed)}$ is the
pulsed luminosity in the 2~--~10~keV band for 1~sr beaming, and
$\dot{E}_{\rm rot,38}$ is the spin-down luminosity in units of
$10^{38}$~erg~s$^{-1}$.  Based on this relationship, the pulsed
luminosity from \psr\ is predicted to be $\sim 7.5 \times
10^{30}$~erg~s$^{-1}$, which is smaller than our best upper limit.
Thus, the result is not yet strongly constraining.

\acknowledgments

This research was partially supported by NASA through the grant
NAG~5-2730.  MH acknowledges support from Research Fellowships of the
Japan Society for the Promotion of Science for Young Scientists.  LRC
acknowledges support by NASA through the grants NAG~5-2948 and
NAG~5-2032 for analysis of ASCA and OSSE data.  JEG acknowledges
support by NASA contract S-10987-C for OSSE data analysis.

\clearpage

\onecolumn

\plotone{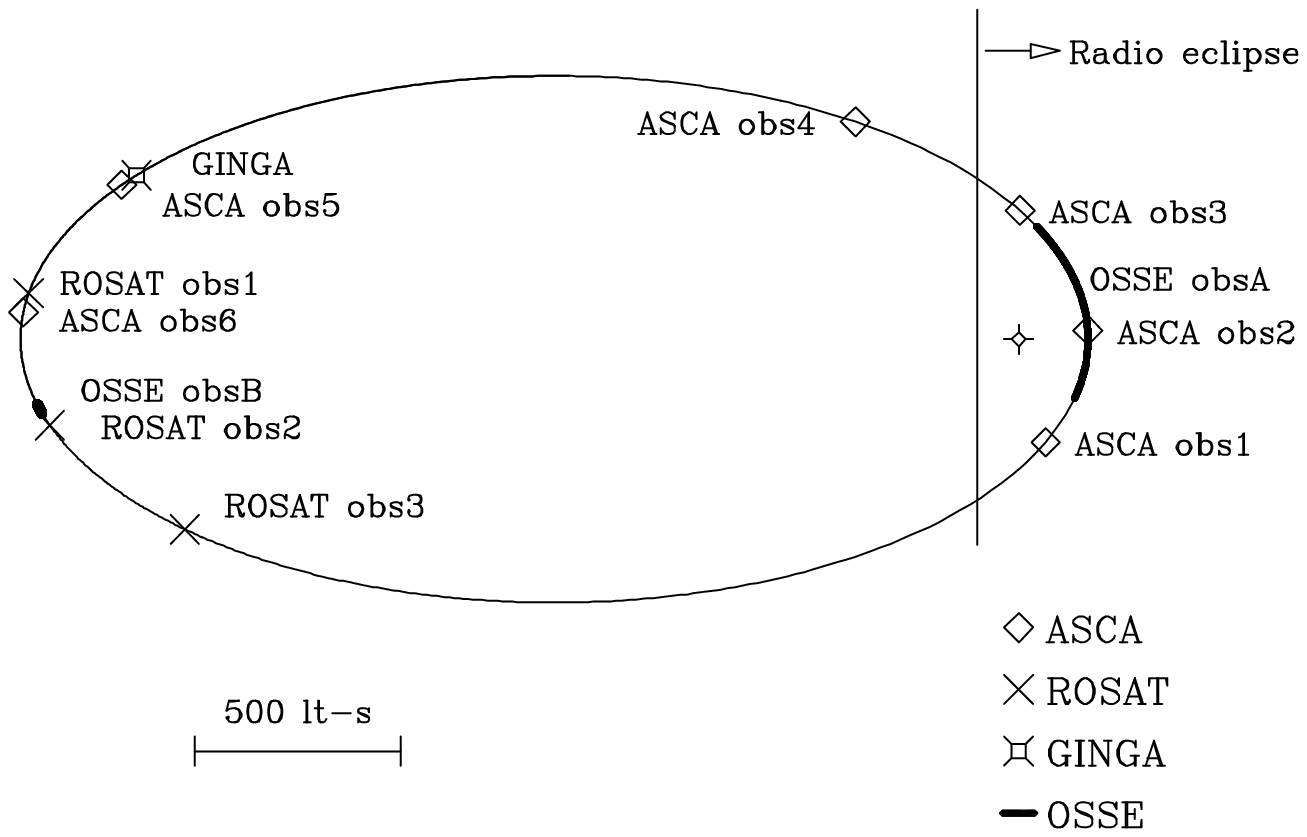}

\figcaption[psr_orbit.ps]{Binary orbit of the \system.  An elliptical
line shows the orbit of \psr\ and a diamond with a cross in the
ellipse indicates the center of gravity of the system.  The figure
also shows the positions of \psr\ when observed with {\it ASCA} (this
work), {\it ROSAT} (Cominsky et al. 1994; Greiner et al. 1995), and
{\it Ginga} (Makino \& Aoki 1994) at various locations on the orbit.
The OSSE instrument on the {\it CGRO} satellite observed \psr\ for
three weeks during the periastron passage in 1994 and for two weeks
near apastron in 1995/6 as indicated with thick lines on the orbit.
When the pulsar is at the right side of the vertical line in the
figure, no radio pulsations are detected (Johnston et al. 1992b).
\label{figure:psr-orbit}}

\clearpage

\plotone{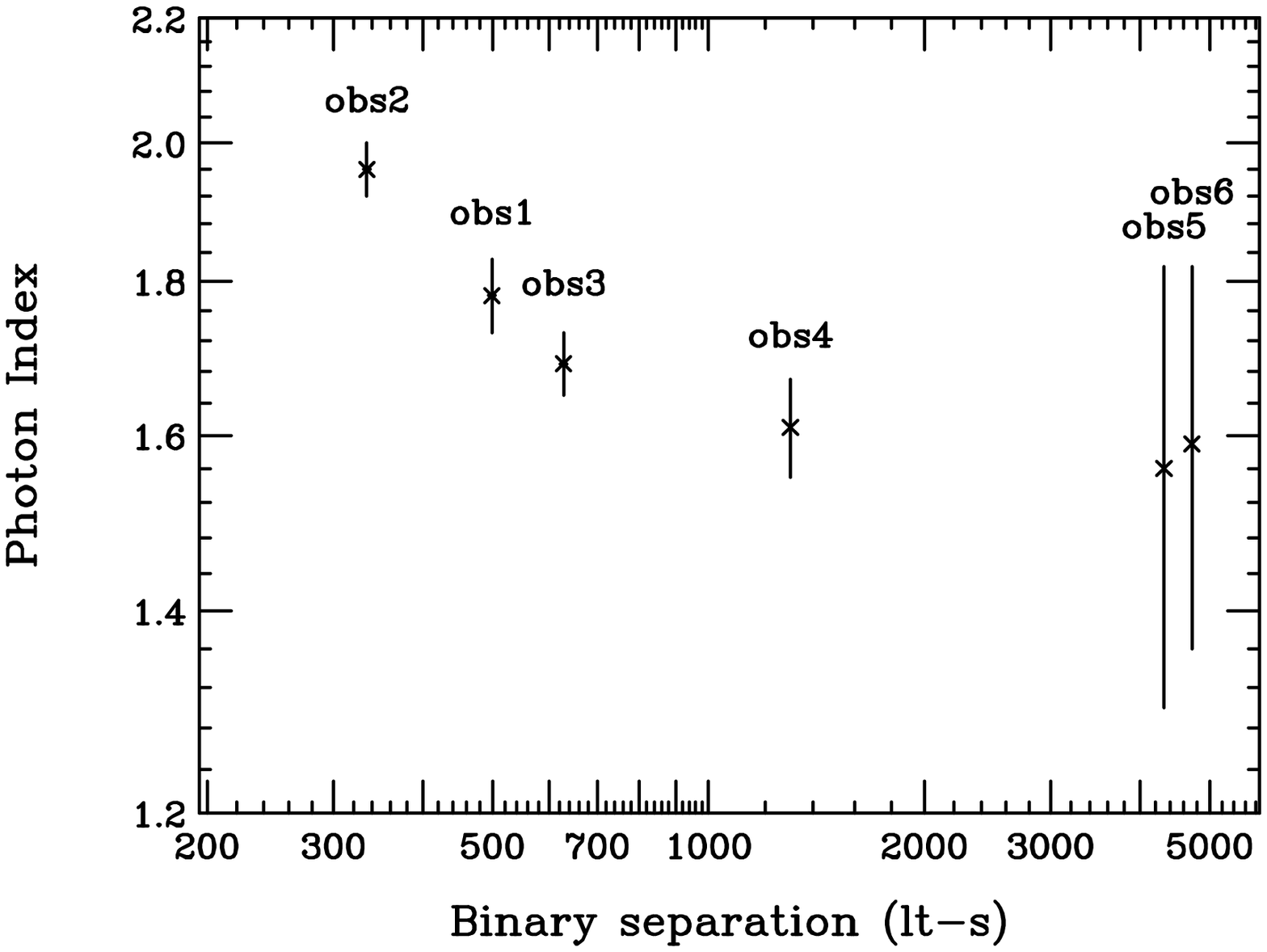}

\figcaption[index_vs_separation.ps]{Photon indices of the \system\ from
{\it ASCA} obs~1 through {\it ASCA} obs~6 are plotted against binary
separation assuming 1.4~\msol\ for \psr\ and 10~\msol\ for SS~2883.
\label{figure:index-vs-separation}}

\clearpage

\plotone{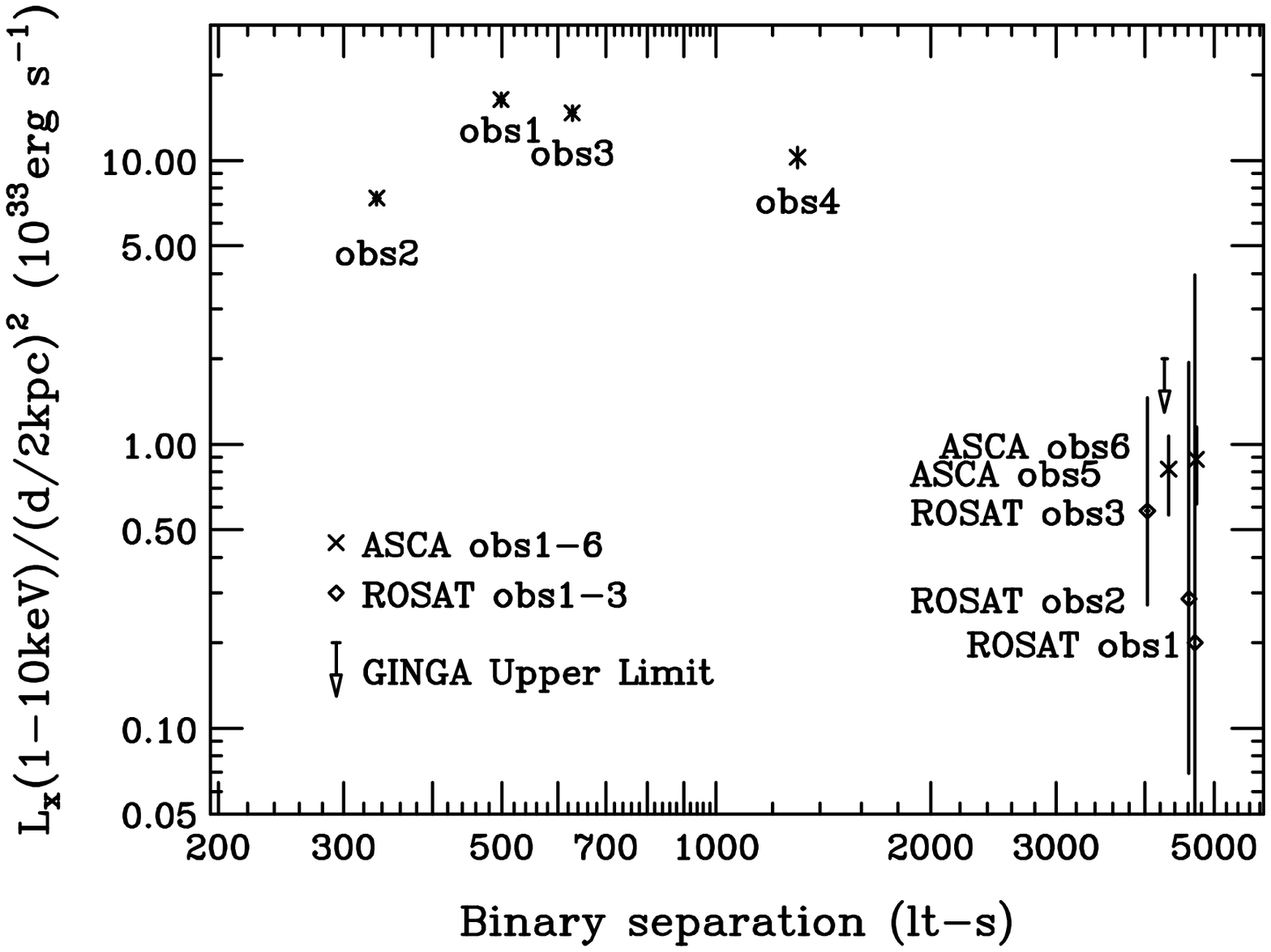}

\figcaption[flux_vs_separation.ps]{Luminosities of the \system\ from
{\it ASCA} obs~1 through {\it ASCA} obs~6 are plotted against a binary
separation assuming masses of 1.4~\msol\ for \psr\ and 10~\msol\ for
SS~2883, and a distance to the \system\ $d=2$~kpc.  In the figure
results from the previous X-ray missions are also plotted (Makino \&
Aoki 1994; Cominsky et al. 1994; Greiner et al. 1995) with the
luminosities in the literature extrapolated to the 1--10~keV band.
\label{figure:flux-vs-separation}}

\clearpage

\plotone{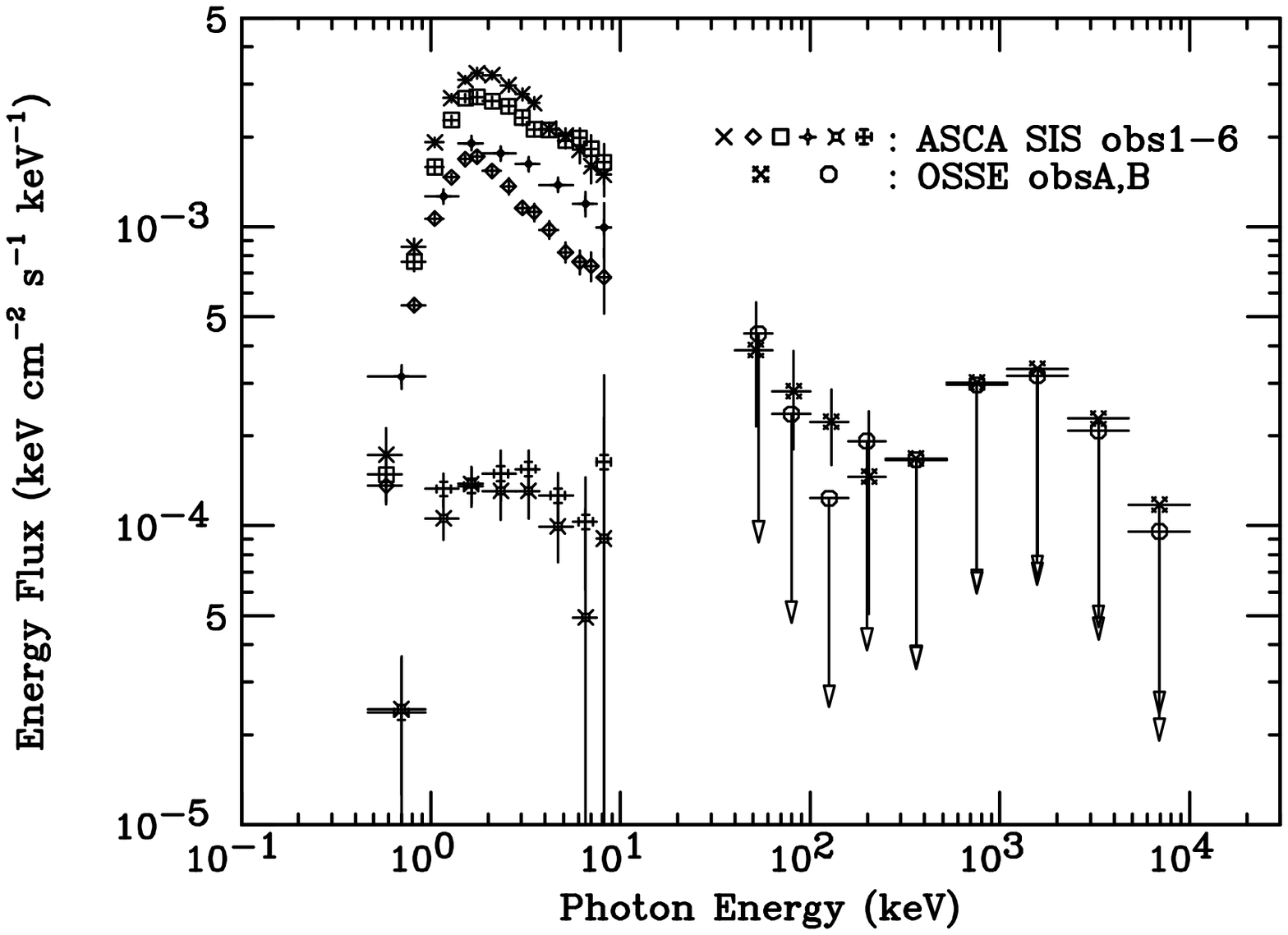}

\figcaption[widespec.ps]{Unfolded spectra obtained for the six {\it
ASCA} observations (Kaspi et al. 1996; Hirayama et al. 1996; this
paper) and the two OSSE observations (Grove et al. 1995; this paper).
For the figure, masses of 1.4~\msol\ for \psr\ and 10~\msol\ for
SS~2883, and a distance to the \system\ $d=2$~kpc are assumed.  Note
that the exposure times for the OSSE observations differ (see
table~\protect\ref{table:observation}).
\label{figure:wideband-spectrum}}

\clearpage

\begin{deluxetable}{rccccl}
\tablecaption{Geometry of the \system\ near the {\it ASCA} and
OSSE observations \label{table:observation}} \tablehead{
\colhead{Observation} & \colhead{MJD} & \colhead{True
Anomaly\tablenotemark{a}} & \colhead{$s
(10^{12}$~cm)\tablenotemark{b}} & \colhead{$s/{\rm R}_{\rm
c}$\tablenotemark{c}} & \colhead{Reference} } \startdata {\it
ASCA} obs~1 & 49349 & $-75^{\circ}$ & 15 & 36 & Kaspi et al. 1995
\nl obs~2 & 49362 & $  8^{\circ}$ & 10 & 24 & Kaspi et al. 1995
\nl obs~3 & 49378 & $ 90^{\circ}$ & 18 & 45 & Kaspi et al. 1995
\nl obs~4 & 49411 & $127^{\circ}$ & 39 & 93 & Hirayama et al. 1996
\nl obs~5 & 49755 & $170^{\circ}$ & 130 & 310 & this paper \nl
obs~6 & 49942 & $178^{\circ}$ & 140 & 340 & this paper \nl OSSE
obs~A & 49355 -- 49375 & $-47^{\circ} \sim +81^{\circ}$
    & 10 -- 16 & 24 -- 39 & Grove et al. 1995 \nl
obs~B & 50071 -- 50084 & $-176^{\circ}$ & 140 & 340 &
this paper \nl \enddata

\tablenotetext{a}{True anomaly is zero at periastron.}
\tablenotetext{b}{Binary separation for assumed Be star and pulsar
masses of $M_c$=10~\msol\ ~and $M_p$=1.4~\msol.}
\tablenotetext{c}{Assuming a companion radius $R_c$ of $6\rsol$.}
\end{deluxetable}

\begin{deluxetable}{rccccc}
\tablecaption{Model parameters for the {\it ASCA} observations of the \system\
\label{table:simultaneous-fit}}
\tablehead{
    \colhead{\it ASCA}
    & \colhead{$N_{\rm H}$\tablenotemark{a}}
    & \colhead{Photon}
    & \colhead{1--10~keV Flux\tablenotemark{a}}
    & \colhead{}
    & \colhead{Fe Emission-Line Flux\tablenotemark{b}}\\
    \colhead{Dataset}
    & \colhead{(10$^{22}$ cm$^{-2}$)}
    & \colhead{Index\tablenotemark{a}}
    & \colhead{(10$^{-11}$~erg~cm$^{-2}$ ${\rm s}^{-1}$)}
    & \colhead{$\chi^2_\nu$}
    & \colhead{($10^{-5}$~photons~cm$^{-2}$~s$^{-1}$)}
}
\startdata
obs~1 & 0.60(4) & 1.78(5) & 3.43(19)/2.96(16) & 0.97 & $<$ 1.7\nl
obs~2 & 0.58(3) & 1.96(4) & 1.54(8)/1.42(7)   & 0.96 & $<$ 1.5\nl
obs~3 & 0.58(4) & 1.69(4) & 3.08(18)/2.76(16) & 1.22 & $<$ 4.0\nl
obs~4 & 0.56(6) & 1.61(6) & 2.15(18)/1.88(16) & 0.98 & $<$ 2.9\nl
obs~5 & 0.5(3)  & 1.6(3)  & 0.17(5)/0.18(5)   & 0.71 & $<$ 0.7\nl
obs~6 & 0.5(2)  & 1.6(2)  & 0.19(6)/0.18(5)   & 0.84 & $<$ 1.3\nl
\enddata
\tablenotetext{a}{Numbers in parentheses represent
90\% confidence interval uncertainties in the last digit quoted.
The uncertainties quoted are statistical,
and do not include any contribution
for unknown systematic calibration errors.}
\tablenotetext{b}{Upper limits with 99 \% confidence assuming a narrow
emission line (of equivalent width of 1~eV) at 6.4~keV, 6.7~keV, or 6.9~keV.}
\end{deluxetable}

\begin{deluxetable}{rcccccc}
\tablecaption{Upper limits on X-ray pulsations obtained from the
{\it ASCA} observations\tablenotemark{a}.
\label{table:upper-limit}} \tablehead{ \colhead{\it ASCA} &
\colhead{} & \colhead{} & \colhead{}\\ \colhead{Dataset} &
\colhead{0.5--10~keV} & \colhead{0.5--2~keV} & \colhead{2--10~keV}
} \startdata obs~1   & 8.71\% (0.0338)
    & 15.1\% (0.0204)
    & 11.8\% (0.0280)\nl
obs~2   & 7.58\% (0.0155)
    & 13.4\% (0.0114)
    & 10.7\% (0.0125)\nl
obs~3   & 8.38\% (0.0302)
    & 18.4\% (0.0243)
    & 11.3\% (0.0253)\nl
obs~4   & 15.1\% (0.0378)
    & 22.2\% (0.0187)
    & 15.3\% (0.0249)\nl
obs~5   & 29.1\% (0.0110)
    & 44.8\% (0.00642)
    & 36.3\% (0.00828)\nl
obs~6   & 29.9\% (0.0125)
    & 49.6\% (0.00783)
    & 37.0\% (0.00902)\nl
\enddata

\tablenotetext{a}{Upper limits in the table are estimated with
99\% confidence, being given as the fractional amplitude of an
assumed sinusoidal pulse.  Counting rates corresponding to the
upper limits are also listed in parentheses in the table in units
of photons~s$^{-1}$.  Note that, due to the smaller luminosities
in {\it ASCA} obs~5 and {\it ASCA} obs~6, these data give higher
fractional upper limits, but stricter upper limits for the pulsed
flux.}
\end{deluxetable}

\end{document}